\documentclass[prl,twocolumn,english,showpacs]{revtex4-1}
\usepackage{graphicx}
\usepackage{bbold}
\usepackage{bm}
\usepackage{amsmath}
\usepackage{amsfonts}
\usepackage{amssymb}
\usepackage{braket}

\begin{document}

\title{Cavity-induced back-action in Purcell-enhanced photoemission of a single ion in an ultraviolet fiber-cavity}

\author{T. G. Ballance$^{1,2}$, H. M. Meyer$^1$, P. Kobel$^1$, K. Ott$^{3}$, J. Reichel$^{3}$ and M. K{\"o}hl$^{1}$}

\affiliation{$^1$Physikalisches Institut, University of Bonn, Wegelerstrasse 8, 53115 Bonn, Germany\\
$^2$Cavendish Laboratory, University of Cambridge, JJ Thomson Avenue, Cambridge CB3 0HE, United Kingdom\\$^3$ Laboratoire Kastler Brossel, \'Ecole Normale Sup\'erieure, 24 Rue Lhomond, 75005 Paris, France}

\begin{abstract}
We study the behavior of a single laser-driven trapped ion inside a microscopic optical Fabry-Perot cavity. In particular, we demonstrate a fiber Fabry-Perot cavity operating on the principal $S_{1/2}\to P_{1/2}$ electric dipole transition of an Yb$^+$ ion at 369\,nm with an atom-ion coupling strength of $g=2\pi\times 67(1)$\,MHz. We employ the cavity to study the generation of single photons and observe cavity-induced back-action in the Purcell-enhanced emission of photons. Tuning of the amplitude and phase of the back-action allows us to enhance or suppress the total rate of photoemission from the ion-cavity system.
\end{abstract}

\maketitle

In light-matter interaction, the Purcell effect has been one of the early fundamental effects studied \cite{Purcell1946,Goy1983} and it has provided the seed for the now well-established field of cavity quantum-electrodynamics. However, despite the long and rich history, there are still principle questions which have not yet been addressed experimentally, for example, the role of cavity-induced back-action \cite{Alsing1992}. Moreover, in recent years, research on the Purcell effect has revived because of its functionality for single photon sources and quantum network nodes in a variety of experimental platforms \cite{Heinzen1987,Kreuter2004,Steiner2013,Lohdahl2004}. A key parameter quantifying the Purcell effect is the  coupling strength $g$ between a single emitter and a single mode of the cavity.  It depends on the electric dipole moment $d$ of the atomic transition and the mode volume $V$ of the cavity, $g \propto d/\sqrt{V}$. In order to obtain near-unity efficiency of photon emission and absorption from a single emitter, it is desirable  to work with small mode volumes and transitions exhibiting strong dipole moments.

The arguably most advanced stationary quantum bits, trapped atomic ions,have been investigated as controlled sources of single photons  \cite{Blinov2004,Maunz2007,Moehring2007,Kurz2014,Higginbottom2016} and in cavity-QED settings \cite{Guthohrlein2001,Stute2012,Steiner2013}, however, yet with relatively small values of $g$. In recent years,  optical cavities for trapped ions have been miniaturized to an ion-mirror separation of less than 100\,$\mu$m \cite{Steiner2013,Steiner2014,Meyer2015} by the advent of optical fiber cavity technology \cite{Hunger2010,Brandstaetter2013,Takahashi2013}. However, further miniaturization of the cavity would be complicated by the high sensitivity of ions to dielectric surfaces~\cite{Harlander2010}.  A remaining tuning knob to enhance light-matter interaction strength for trapped ions is the electric dipole moment $d$. So far, fiber cavities for trapped ions have operated in the near-infrared spectral range on relatively weak transitions between metastable states \cite{Steiner2013,Steiner2014,Meyer2015}. This was motivated by the availability of well-proven fiber, coating, and laser machining technology in this spectral range. However, the principal optical dipole transitions of trapped ions are in the ultraviolet and blue range of the optical spectrum, a regime traditionally challenging for optical fiber technology. Previous work on optical cavities in the ultraviolet and blue regime has focussed on macroscopic, mm-long cavities using conventional mirrors \cite{Guthohrlein2001,Leibrandt2009,Sterk2012,Bylinskii2015}. In this paper we demonstrate a fiber Fabry-Perot cavity resonant with the $S_{1/2} \to P_{1/2}$ transition of singly-charged ytterbium ions at 369\,nm (see Fig. 1). Owing to the smaller mode volume and higher dipole moment as compared to  previous  experiments, we boost the atom-photon coupling strength by more than one order of magnitude. We employ the cavity to study the generation of single photons on this transition and we show back-action of the generated intracavity field onto the ion's emission rate, which can be tuned to enhance or suppress the total emission rate.

\begin{figure}
\includegraphics[width=\columnwidth]{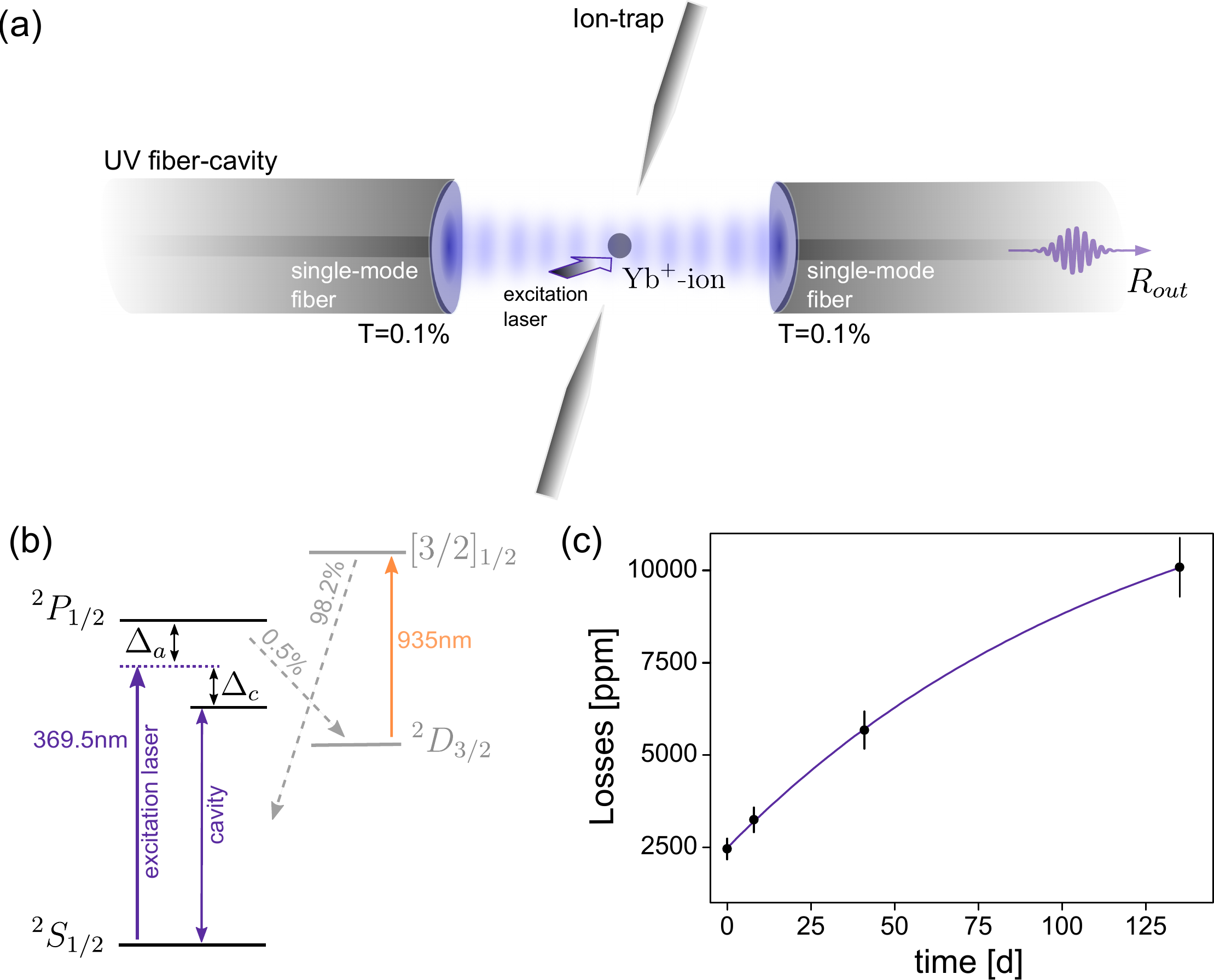}
\caption{(Color online) (a) Schematic setup comprising of an endcap trap and the fiber cavity. (b) $^{174}$Yb$^+$ ion level structure. (c) Loss increase of the cavity coating vs. time under ultrahigh vacuum conditions measured from the decrease of the cavity in-coupling efficiency. The increase fits the exponential model described in \cite{Gangloff2015} with a time constant of $\tau_{\text{loss}}=123\pm15$ days.}
\label{fig1}
\end{figure}

A single $^{174}\text{Yb}^+$ ion is trapped in a miniaturized radio-frequency Paul trap with integrated fiber-cavity (Fig. \ref{fig1}a). Applying a radio-frequency signal (amplitude $75\,$V, frequency $30\,$MHz) to the two opposing needle-shaped endcap electrodes~\cite{Steiner2013,Deslauriers2006}, which are separated by $84\,\mu$m, leads to a harmonic trapping potential with trap frequencies of 2-5~MHz. Laser cooling and fluorescence detection of the ion are performed on the $^2$S$_{1/2}$--$^2$P$_{1/2}$ transition near 369\,nm (Fig.~\ref{fig1}b). Spontaneous decay from the excited $^2$P$_{1/2}$ state populates the metastable $^2D_{3/2}$ state with a branching ratio of 1/200. The metastable $^2D_{3/2}$ state is cleared out by a repumping laser at 935\,nm.
Our cavity consists of a pair of single-mode optical fibers, which have been laser-machined to feature near-spherical surfaces of $\sim 200\, \mu$m radius of curvature \cite{Hunger2010}. The fiber tips have been coated with a high-reflectivity coating with a transmission of $T=1000$\,ppm at a wavelength of $369$\,nm and losses are on the order of 500\,ppm.
A pair of fiber tips form a Fabry-Perot cavity with a length of $150\,\mu$m. Although the cavity design parameters correspond to a finesse of $F=2000$, we have observed an initial cavity finesse of $F=1140$, which degraded to the final value of $F=209$ over the course of 3 months under ultrahigh vacuum conditions by an as yet unknown process (see Fig. 1c). Our high-reflectivity coating features a top layer of SiO$_2$ with a thickness of $120$\,nm, which is expected to prevent degradation from oxygen diffusion \cite{Gangloff2015}. Part of the finesse degradation was recovered by exposing the cavity mirrors to air and hence we exclude contamination from ytterbium deposition. Finesse and length result in a field decay rate of  $\kappa= 2\pi \times 2.4(1)$\,GHz. The fundamental Gaussian resonator mode exhibits a mode waist ($1/e^2$--radius of the intensity profile) of $w=(3.1 \pm 0.1)\,\mu$m, which gives rise to a theoretical atom-field coupling strength of  $g=2\pi\times 96$\,MHz for circular polarization in the $J=1/2 \to J^\prime=1/2$ system. This constitutes by more than an order of magnitude the highest coupling strength between a single trapped ion and a single-mode of the radiation field and will prove of high relevance for future applications as quantum network node. The decay rate of the atomic dipole moment of the $P_{1/2}$ excited state is $\gamma=\Gamma/2=2\pi \times 9.8$\,MHz, which places the cavity at a cooperativity $C_0=g^2/(2\kappa \gamma)=0.2$. 

First, we demonstrate that the single ion in the cavity acts as single-photon source employing Purcell enhancement of the excited state decay rate. The Purcell effect generally leads to an enhanced excited-state decay rate of $\Gamma'=\Gamma(2C_0+1)$ when the cavity is resonant with the atom and only covers a negligible fraction of the total solid angle. The so-called Purcell factor $f_P$ describes the ratio between emission rate into the cavity mode $P_c$ and the emission rate into all free-space modes $P_{4 \pi-c}$. It is directly linked to the cooperativity $C_0$ by $f_P=P_c/P_{4\pi-c}= 2C_0 \kappa^2/(\kappa^2+\Delta_{ac}^2)$, where $\Delta_{ac}=\omega_c-\omega_a$ is the detuning between the cavity ($\omega_c$) and the atomic ($\omega_a$) resonance frequency.  In the experiment we drive the ion with a near-resonant laser field of frequency $\omega$ applied transversely to the cavity and we collect the photons emitted from the ion into the cavity mode at the output of the fiber. The detuning of the atomic  and the cavity resonance frequency with respect to the drive laser frequency is defined as $\Delta_a=\omega-\omega_a$ and $\Delta_c=\omega-\omega_c$, respectively. For single-photon generation, we choose the cavity resonance frequency $\omega_c$ equal to the atomic resonance frequency $\omega_a$, and the drive laser frequency is tuned one atomic linewidth $\Gamma$ to the red. For a saturation parameter of $s= 14$, we achieve a photon production rate out of the single-mode fiber of $R_{out} = 2.1(4) \times 10^5\, \text{s}^{-1}$. 

\begin{figure}
\includegraphics[width=0.85\columnwidth]{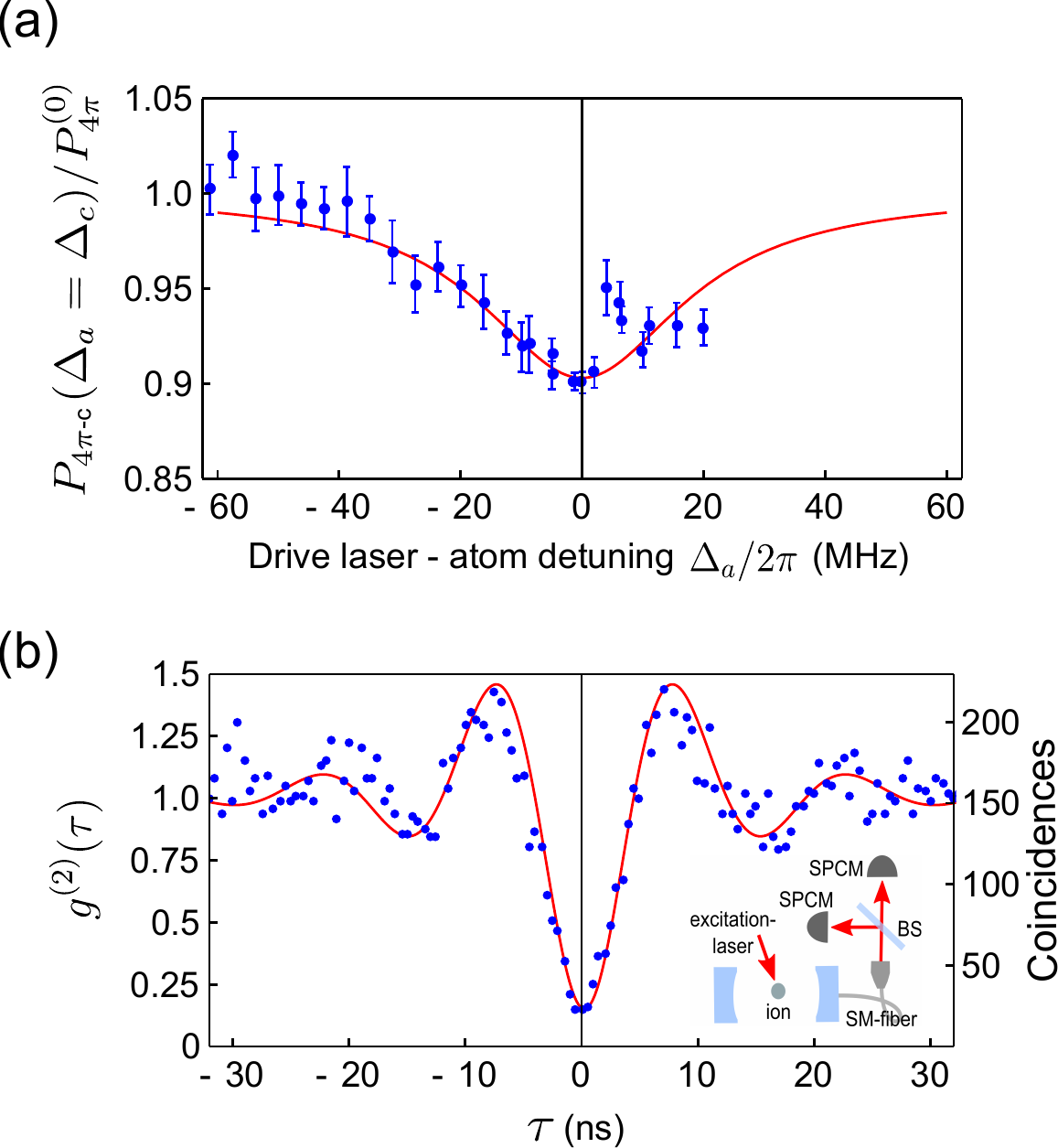}
\caption{(Color online)  (a) Photon rate scattered transversely to the cavity mode vs. atom-laser detuning. The solid line is a Lorentzian fit to the data from which the Purcell enhanced linewidth of the atomic transition can be retrieved (see main text). The larger scattering of the data at positive detuning comes from Doppler heating of the ion during the measurement cycle. The saturation parameter for this measurement is $s=2.8$, from which follows $\Gamma'=2\pi \times 24(5)$. (b) Second-order correlation function of the photons emitted from the cavity tuned into resonance with the ion ($\Delta_a=\Delta_c$) and a drive laser detuning of $\Delta_a = -\Gamma/2$ at $s=14$. The solid line is a fit to the data (see text). Inset: Experimental setup. BS: beam-splitter, SPCM: single-photon counting module, SM: single-mode.}
\label{fig2}
\end{figure}

In order to analyze the photoemission quantitatively, we first characterize the increased linewidth $\Gamma'$ resulting from the Purcell effect. To do so, we monitor the drop of the fluorescence rate of the ion collected transversely to the cavity as we change the atom-laser detuning $\Delta_a$ while keeping the cavity resonant with the atomic transition ($\Delta_c=\Delta_a$) and normalize it to the case without cavity $P_{4\pi}^{(0)}$. In our parameter regime, this curve is well described by a Lorentzian profile of full width $\sqrt{\Gamma'^2+s\cdot\Gamma^2}=\Gamma\sqrt{(2C_0+1)^2 + s}$, which can be interpreted as a Purcell- and saturation broadened resonance, providing us with the fit result $\Gamma^\prime=2\pi \times 24(5)$~MHz (see Fig. 2a). 
Next, we employ a Hanbury Brown and Twiss setup of two single-photon counters and a non-polarizing beam splitter and confirm that the emitted photons exhibit anti-bunching as shown in Fig. \ref{fig2}b. We fit our correlation data with the theoretical intensity correlation function of a driven two-level system \cite{Dagenais1978} using the measured Purcell-enhanced linewidth  $\Gamma^\prime$. In order to account for the measured jitter of the single-photon counters we convolve the ideal correlation function with a Gaussian of full width at half maximum of 3.2\,ns. The finite value of the correlation function of $g^{(2)}(0)=0.15$ is almost exclusively accounted for by timing jitter of the detectors whereas detector dark counts contribute only $\sim 2\cdot10^{-2}$, which cannot be resolved from our measurement. 

\begin{figure*}
\includegraphics[width=2\columnwidth]{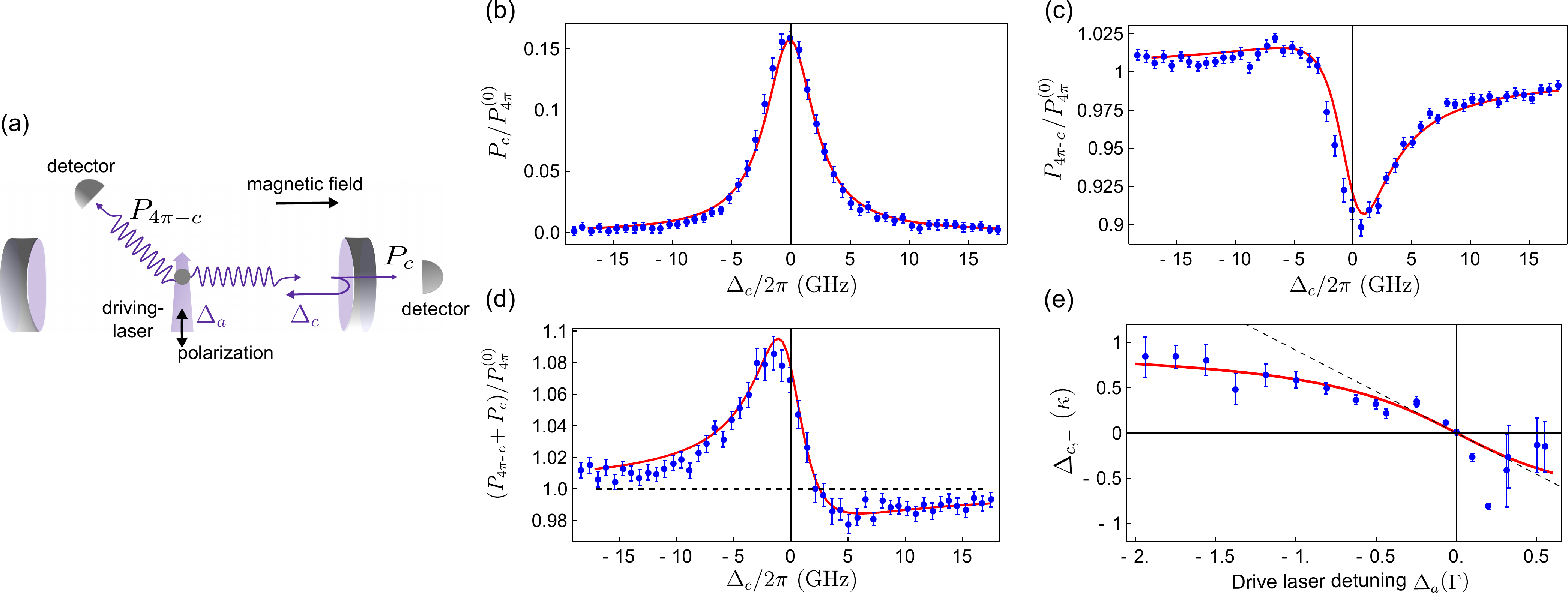}
\caption{(Color online) (a) Two-level system in a cavity driven by a laser transverse to the cavity. Photodetectors monitor both rate $P_c$ and $P_{4\pi-c}$. A magnetic field of about 2G is applied along the cavity axis such that the cavity only mediates $\sigma$-transitions and the drive laser polarization is adjusted perpendicular to it. (b) Cavity output. (c) Free-space emission.  (d) Total photoemission. (e) Minimum of the free-space scattering $\Delta_{c,-}$ for different detunings of the drive laser from the atomic resonance. The solid line is from eqn. \ref{eqn:cmins} and the dashed line describes the linear dependance of the minimum for a drive laser detuning close to atomic resonance discussed in the main text.}
\label{fig3}
\end{figure*}

However, the physics of the atom-cavity interaction is more involved than the simple Purcell-enhanced photoemission picture. In particular, part of the light field generated by the atom is stored in the cavity  and leads to cavity-induced back-action onto the emitter. Cavity-induced back-action arises in the following way: upon driving the ion with a laser transverse to the cavity mode with an electric field $E_d e^{-i\omega t}$, the ion responds according to its electric polarizability $\alpha(\omega)$ and scatters an electric field $\propto \alpha(\omega) E_d$ both into the cavity mode and into vacuum. The latter can be considered coupling to a Markovian environment since the photons never  return to the ion. However, the electric field scattered into the cavity mode is reflected from the mirrors with a certain amplitude and leads to an additional electric field at the ion. The intracavity electric field  $E_c$ and the driving electric field $E_d$ interfere with each other at the location of the ion, leading to a modification of the excitation rate of the ion. As an extreme case, it has been theoretically pointed out \cite{Alsing1992} that resonance fluorescence of an emitter inside a cavity can be completely suppressed in the case of a lossless cavity  owing to fully destructive interference between the driving field and the field building up in the cavity. Moreover, the interference of light emitted from a  single ion with its mirror image has been observed without a cavity \cite{Eschner2001}. Generally,  Purcell's argument regarding the ratio between emission rates into the cavity-mode and free-space modes  remains quantitatively valid, however, depending on the degree of back-action, the {\it total} emission rate of the atom in the cavity $P_c+P_{4\pi-c}$ can be enhanced or suppressed relative to the case without cavity $P_{4\pi}^{(0)}$, an effect which has not yet been observed.

Let us put the effect of the back-action on more precise grounds, which we can do analytically in the weakly driven case where saturation of the atom can be neglected \cite{TanjiSuzuki2011}: The generated intra-cavity field has the same frequency as the driving field but the phase relative to the driving field is set by the detunings $\Delta_a$ and $\Delta_c$. The detuning $\Delta_a$ controls the relative phase between the driving field and the field radiated by the atomic dipole, and $\Delta_c$ tunes the phase accumulated in the round trip through the cavity. In the near-resonant regime, $\Delta_c \ll \kappa$, $\Delta_a \ll \Gamma$ and for weak cooperativity $C_0 <1$, we can approximate the intra-cavity field by $
\frac{E_c}{E_d}\simeq -C_0 \frac{\Gamma^2}{\Gamma^2+ \Delta_a^2}\exp\left[i\left(\frac{\Delta_c}{\kappa}-\frac{\Delta_a}{\Gamma}\right)\right]$. The exponential factor contains the relevant physics: it measures the difference between the phase shift of a photon in a detuned resonator of width $\kappa$ and the photon emitted from a driven dipole. When these two phase shifts cancel each other, the intracavity electric field is exactly $\pi$ out of phase with the drive laser field and the excitation of the ion is reduced by the destructive interference.  However, the destructive interference is only partial since the amplitude of the intracavity field is diminished by the Lorentzian prefactor with detuning $\Delta_a$. Therefore, our setup allows for independently tuning the phase and the amplitude of the back-action of the cavity environment onto the ion. The principal argument given here is valid for all detunings, and constructive (+) and destructive (--) interference occur for 
\begin{equation}
\frac{\Delta_{c,\pm}}{\kappa}=\frac{\Gamma  \left(C_0+1\right)\pm\sqrt{ 4 \Delta_a^2+\Gamma ^2 \left(C_0+1\right)^2}}{2 \Delta_a}.
\label{eqn:cmins}
\end{equation}

A recent related paper investigating neutral atoms in a highly-dissipative cavity \cite{Lien2016} employs a quantum mechanical description to derive the same physics in an EIT-related language. However, the classical description based on the atomic polarizability and classical fields suffices to quantitatively explain the effects for low excitation powers. For large transverse excitation intensities, a quantum mechanical description based on the solution of the master equation for the driven Jaynes-Cummings model is required to account for saturation of the atomic transition.

In Figure \ref{fig3}, we show the observation of cavity-induced back-action in the Purcell effect. We excite the ion with a laser transverse to the cavity and measure the emission rate both into the cavity $P_c$ (Fig. \ref{fig3}b) and into free-space $P_{4\pi-c}$ (Fig. \ref{fig3}c). The given example is for $\Delta_a = -\Gamma/2$ and a saturation parameter of the drive laser of $s=2.8$. Both observed emission rates are normalized to the free-space emission rate in the absence of the cavity $P_{4\pi}^{(0)}$, which can be directly measured in the case of $\Delta_c\gg\kappa$ as the solid-angle covered by the cavity is small. 

In order to calibrate the emission rate into the cavity, we determine the cavity impedance- ($\eta_{\text{IM}}=0.033$) and mode-matching efficiency ($\eta_{\text{MM}}=0.5$) from the observed cavity in-coupling, the efficiency of the optical path ($\eta_{\text{path}}=0.5$, excluding fiber loss) as well as the quantum efficiency of the photon detector ($\eta_{\text{PMT}}=0.14$). This in total results in a probability of $\eta = 1.16\cdot 10^{-3}$ for a photon in the cavity mode to be detected. 

The cavity output rate is approximately a Lorentzian curve with its  center  shifted away from zero detuning by $\sim C_0 \kappa = 200$~MHz. The peak height of the Lorentzian measures directly the cooperativity as $2C_0/[1+2C_0+2C_0^2]$ for the given detuning. In contrast, the free-space emission rate (Fig. \ref{fig3}c) exhibits an asymmetric Fano-like profile including a maximum and a minimum of the emission rate at cavity detunings $\Delta_{c,+}$ and $\Delta_{c,-}$, respectively. The sum of the two rates is plotted in Fig. \ref{fig3}d showing both enhancement and suppression of the total emission rate of the ion due to the back-action of the cavity field, depending on the cavity detuning $\Delta_c$. The solid line in Fig. \ref{fig3}b--d are the solution of the master equation of the dissipative Jaynes-Cummings Hamiltonian, where we included a fiber transmission 90\% as a free parameter, which is in agreement with the expected transmission. From this we determine a single-photon coupling rate of $g=2\pi \times 67(1)$\,MHz, corresponding to a cooperativity of $C_0 =0.094(5)$. The deviation from the theoretically expected value of $g=2\pi \times 96$\,MHz is by a similar factor as in our previous experiments in a different setup \cite{Steiner2013,Steiner2014} and reflects an imperfect localization of the ion along the cavity standing-wave field. 

In order to highlight the sensitivity of the cavity-induced back-action to the relative phases between drive laser, atomic dipole and cavity field,  we have measured the free-space emission for different detunings $\Delta_a$ and $\Delta_c$ and extracted the positions of the local minima $\Delta_{c,-}$ of the emission rates using  polynomial fits. The result of this analysis is shown in Fig. \ref{fig3}e and the data agree excellently with the predicted frequencies of destructive interference of equation (\ref{eqn:cmins}). This supports our interpretation as interferences due to cavity back-action. We have not been able to infer the positions of the local maxima $\Delta_{c,+}$ using an independent fit since the curvature near the maxima is very shallow. For $\Delta_a \simeq 0$ we find a linear dependence $\Delta_{c,-}\simeq -\Delta_a\kappa/[\Gamma(1+2C_0)] $ (see Fig. \ref{fig3}e, dashed line). This relation proves very useful as it allows for accurately measuring the cooperativity $C_0$ in the fast-cavity regime since its experimental signature is enhanced by a (large) factor $\sim \kappa/\Gamma$.  Finally, we would like to stress that even if the average occupation number of the cavity mode is only $\bar{n}\sim 2C_0\Gamma/\kappa \sim 10^{-3}$, this still leads to a measurable back-action since the effect is based on the interference of the electric field amplitudes, i.e. proportional to $\sqrt{\bar{n}}\sim 5\%$.

In summary, we have demonstrated an optical fiber cavity in the ultraviolet spectral range coupled to the strong electric dipole transition of a trapped Yb$^+$ ion. We  have demonstrated that the cavity can be used as a tailored environment, which can act to suppress or enhance the total emission rate of an atom by tuning the phase of the back-action field. Our work paves the way towards employing trapped ions on their strong optical-dipole transition in the ultraviolet as nodes in fiber-based quantum networks. 

We would like to thank M. Steiner  for experimental assistance. The work has been supported by BCGS, ITN COMIQ, DFG (SFB/TR 185), and the Alexander-von-Humboldt Stiftung.


%

\end{document}